\documentclass[twocolumn,showpacs,preprintnumbers,amsmath,amssymb,superscriptaddress]{revtex4}

\usepackage{graphicx}
\usepackage{dcolumn}
\usepackage{bm}
\usepackage{hyperref}
\usepackage{latexsym}

\begin{document}

\title{Robust Patterns in Food Web Structure}

\author{Juan Camacho}
\affiliation{Center for Polymer Studies and Dept. of Physics,
             Boston University, Boston, MA 02215} 
\affiliation{Departament de F\'{\i}sica (F\'{\i}sica Estad\'{\i}stica),
Universitat Aut\`onoma de Barcelona, E-08193 Bellaterra, Spain}

\author{Roger Guimer\`a} 
\affiliation{Center for Polymer Studies and Dept. of Physics, Boston
University, Boston, MA 02215} 
\affiliation{Departament d'Enginyeria Qu\'{\i}mica, Universitat Rovira i
Virgili, 43006 Tarragona , Spain}

\author{Lu\'{\i}s A. Nunes \surname{Amaral}}
\affiliation{Center for Polymer Studies and Dept. of Physics,
             Boston University, Boston, MA 02215}   

\begin{abstract}

We analyze the properties of seven community food webs from a
variety of environments---including freshwater, marine-freshwater
interfaces and terrestrial environments.  We uncover quantitative unifying
patterns that describe the properties of the diverse trophic webs considered
and suggest that statistical physics concepts such as scaling and
universality may be useful in the description of ecosystems.  Specifically,
we find that several quantities characterizing these diverse food webs obey
functional forms that are universal across the different environments
considered. The empirical results are in remarkable agreement with the
analytical solution of a recently proposed model for food webs.

\end{abstract}


\maketitle


In natural ecosystems species are connected through trophic relationships
\cite{Martinez00,McKann98,Paine92} defining intricate networks
\cite{Strogatz01,Watts98,Amaral00,Newman01}, the so-called food
webs. Understanding the structure and mechanisms underlying the formation of
these webs is of great importance in ecology \cite{ecology}. For this reason,
much research has been done in constructing empirical webs and uncovering
unifying patterns describing their structure \cite{Lawton88,ecology}.
However, in the last decade the construction of larger and more complete food
webs clearly indicated that the previously reported unifying patterns do not
hold for the new webs \cite{Hall,Hall2}.  Indeed, the complexity of the new
webs has rendered quite difficult the challenge to obtain quantitative
patterns that substitute the old ones.

Here, we analyze the properties of seven detailed community food webs from a
variety of environments---including freshwater habitats, marine-freshwater
interfaces and terrestrial environments.  Remarkably, we uncover quantitative
unifying patterns that describe the properties of the diverse trophic webs
considered and capture the random and non-random aspects of their structure.
Specifically, we find that several quantities---such as the distributions of
number of prey, number of predators, and number of trophic
links---characterizing these diverse food webs obey robust functional forms
that depend on a single parameter, the linkage density $z$.

In our analysis, we use results obtained for complex networks \cite{Newman01}
and for a recent model of food web formation, the ``niche model'' of
Ref.~\cite{Martinez00}. We first describe the theoretical model and its
predictions: An ecosystem with $S$ species and $L$ trophic interactions
between these species, defines a network with $S$ nodes and $L$ directed
links. In the niche model, one first randomly assigns species $i=1,\dots ,S$
to ``trophic niches'' $n_i$ which are mapped into the interval [0,1]. A
species $i$ is characterized by its niche parameter $n_i$ and by its list of
prey. Prey are chosen according to the following procedure: species $i$ preys
on the species $j$ with niche parameters $n_j$ inside a segment of length
$xn_i$ centered in a position chosen randomly inside the interval
$[xn_i/2,n_i]$. Here, $0\le x\le 1$ is a random variable with probability
density function
%
$p_x(x)=b\left( 1-x\right) ^{\left( b-1\right)}$ 
%
\cite{Martinez00}.  The values of the parameters $b\equiv (S^2/2L-1)$ and $S$
determine the linkage density $z\equiv L/S$ of the food web, and the directed
connectance $L/S^2$, which is a measure of the fraction of the actual number
of trophic links as compared to the maximum possible number
\cite{Martinez00}. 


In the limit of large web sizes ($S \gg 1$) and small connectances ($L/S^2
\ll 1$), one can derive analytical expressions for the distribution of number
of prey $k$ \cite{Camacho01}.  We consider the cumulative distribution
$P_{{\rm prey}}(k) = \sum_{k^{\prime} \ge k} p_{{\rm prey}}(k^{\prime})$
because it is less noisy than the probability function $p_{\rm prey}(k)$.  We
obtain
%
%
%
\begin{equation}
P_{{\rm prey}}(k) = \exp\left( -\frac{k}{2z} \right) - \frac{k}{2z}%
\,E_1\left(\frac{k}{2z} \right) \,,  \label{prey}
\end{equation}
where $E_1(x)$ is the so-called exponential-integral function \cite{numrec}.
Equation~(\ref{prey}) predicts that the distribution of number of prey
decays exponentially for large $k$.

Also in the limit of large web sizes and small connectances, one can derive
analytical expressions for the distribution of number of predators $m$
\cite{Camacho01}. We obtain,
%
%
\begin{equation}
P_{{\rm pred}}(m) = \frac{1}{2z} ~\sum_{m^{\prime}=m}^\infty ~\gamma
(m^{\prime}+1,2z) \,,  \label{pred}
\end{equation}
where $\gamma (m+1,z)$ 
is the so-called ``incomplete gamma function'' \cite{numrec}. To gain
intuition about the functional form (\ref{pred}), note that $p_{\rm pred}(m)$
is approximately a step function: It is constant for $m < z$, and then it
decays with a Gaussian tail for $m\approx 2z$ \cite{Camacho01}.
It follows then that the cumulative distribution $P_{\rm pred}(m)$ decreases
linearly as $1- m/z $ for $m < z$ and decays as the error function
\cite{numrec} for $m \approx 2z$.

%
\begin{figure}[t!]
 \vspace*{0.cm}
 \includegraphics*[width=7.5cm]{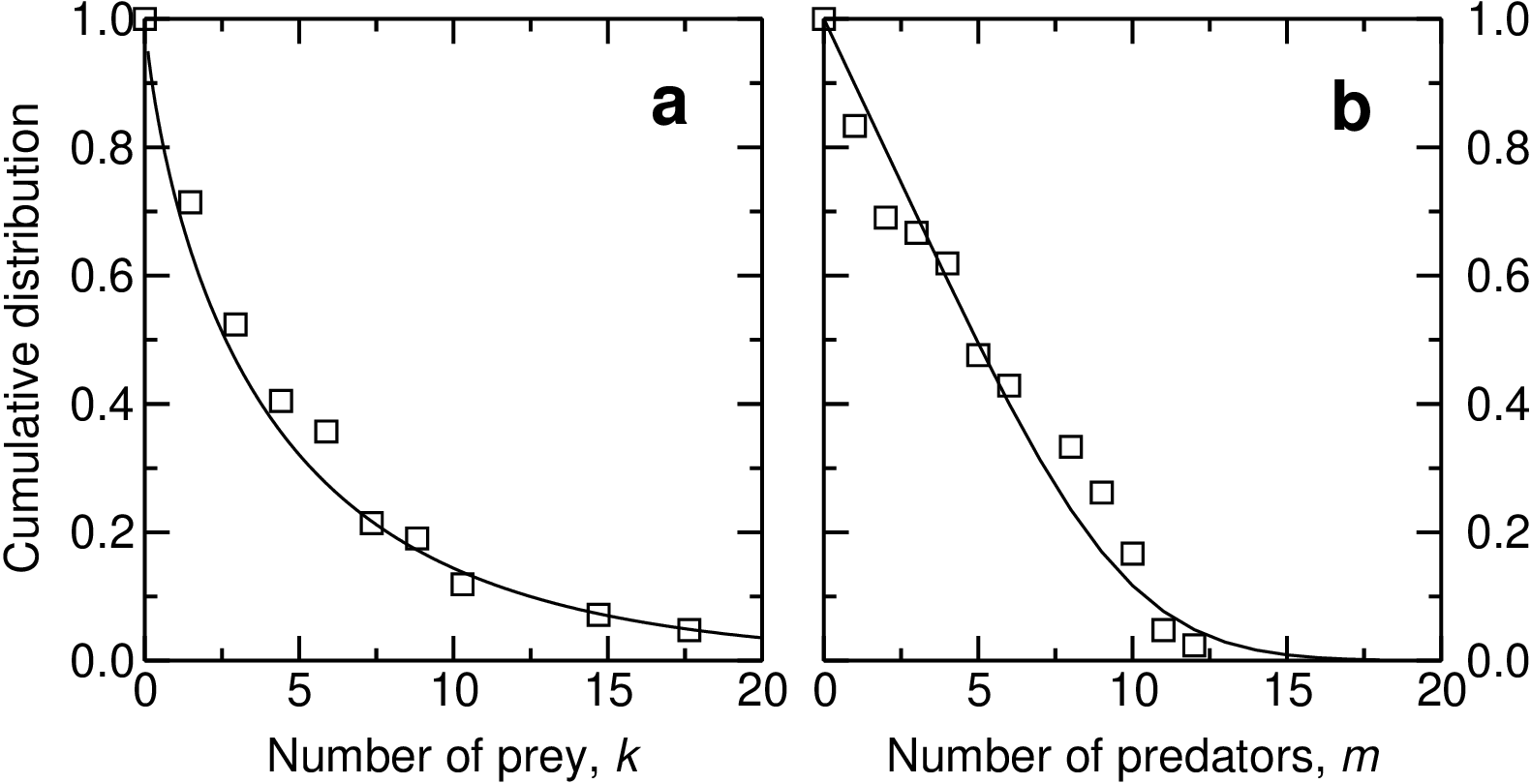}
 \vspace*{-0.3cm}
 \caption{ Cumulative distribution (a) $P_{{\rm prey}}$ of number of prey
  $k$, and (b) $P_{{\rm pred}}$ of number of predators $m$ for the St.\
  Martin Island web. The data agrees well with the analytical predictions of
  Eqs.~(\ref{prey})--(\ref{pred})---indicated by the solid lines, without any
  free parameters for fitting as $z$ is determined empirically. }
 \label{figure1}
\end{figure}
%


Next, we analyze the empirical data for seven food webs with 25 to 92 trophic
species.  These webs have linkage densities $2.2 < z < 10.8$, and
connectances in the interval 0.06--0.31 \cite {Martinez00}. We first
investigate the distributions of number of prey and number of predators. 
Figures~\ref{figure1}a,b compare the cumulative distributions of the number
of prey and number of predators for species in the St.\ Martin Island web
\cite{smartin} with our analytical predictions, and suggest that these
distributions are well approximated by Eqs.~(\ref{prey})--(\ref{pred})
without any free parameters for fitting. Equations~(\ref{prey})--(\ref{pred})
and the results of Fig.~\ref{figure1} suggest the possibility that $P_{\rm
prey}$ and $P_{\rm pred}$ obey universal functional forms that depend only on
$z$.

\begin{figure}[t!]
 \vspace*{0.cm}
 \includegraphics*[width=7.5cm]{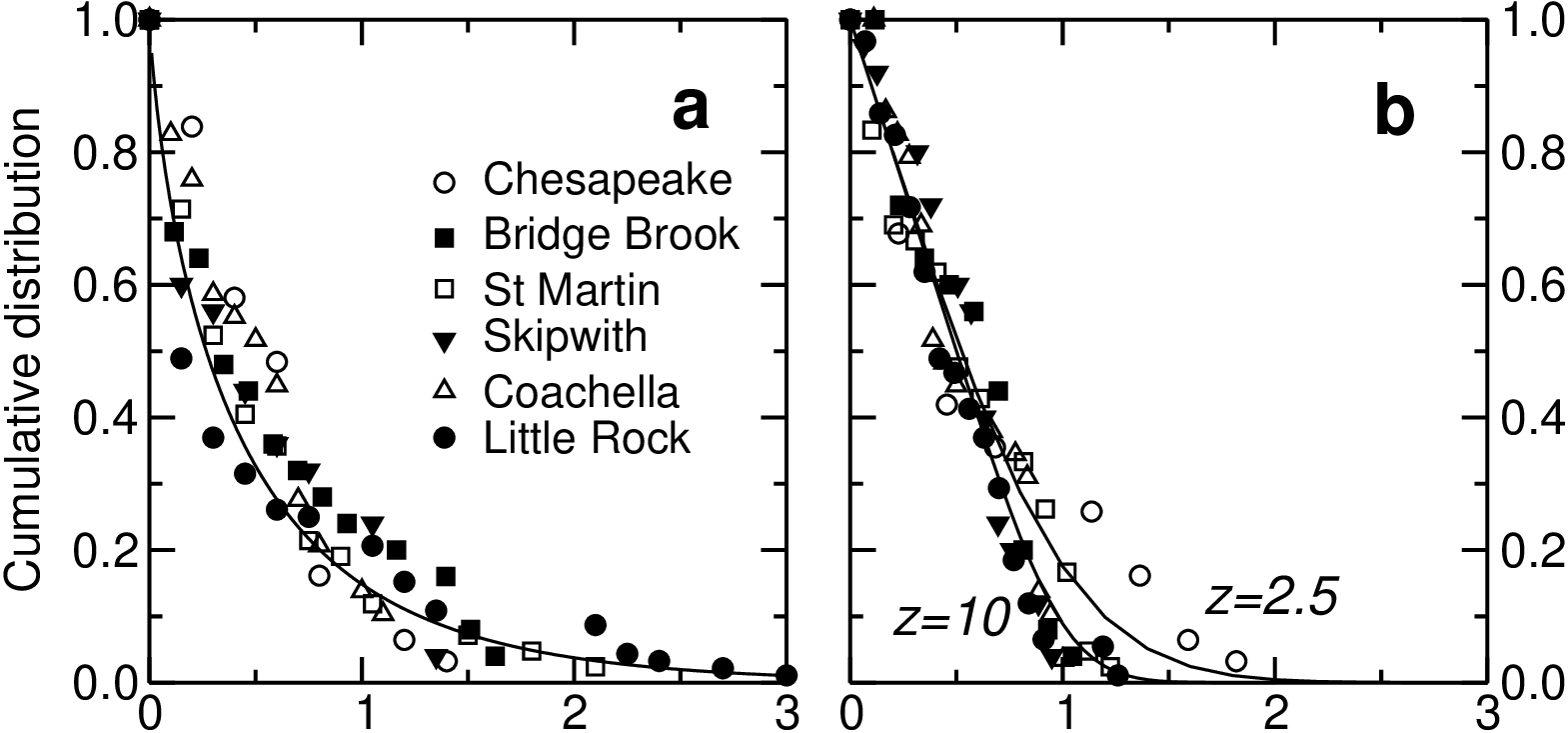}
 \vspace*{0.1cm}
 \includegraphics*[width=7.5cm]{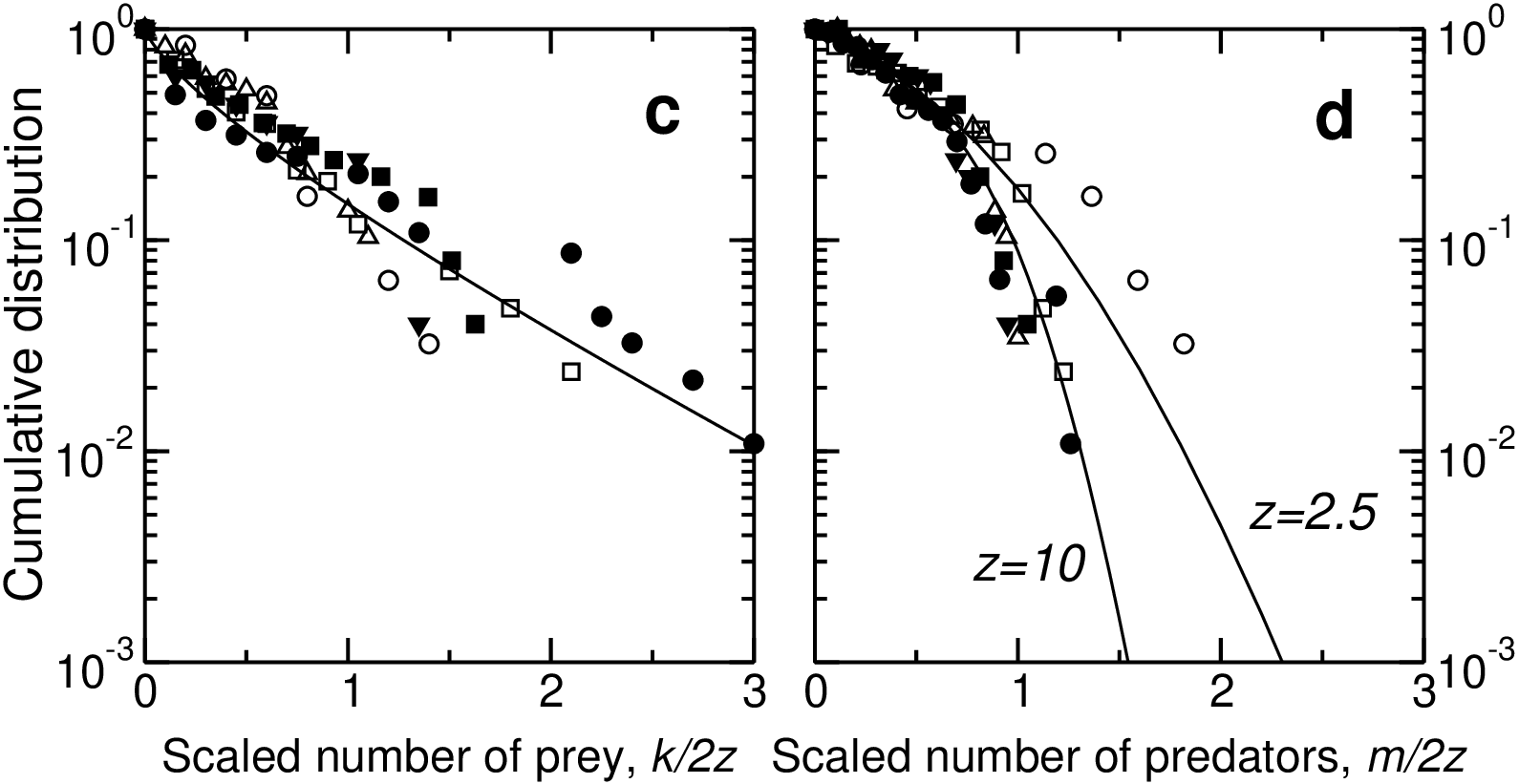}
 \vspace*{-0.3cm}
 \caption{ We test the ``scaling hypothesis'' \protect\cite{Bunde94} that the
  distributions of number of prey (predators) have the same functional form
  for different food webs. (a) Cumulative distribution $P_{\rm prey}$ of the
  scaled number of prey $k/2z$ for all the webs except Ythan
  \protect\cite{ks-test1,ks-test2}.  The solid line is the prediction of
  Eq.~(\ref{prey}).  The data ``collapses'' onto a single curve that agrees
  well with the analytical results. (b) Cumulative distribution $P_{\rm
  pred}$ of the scaled number of predators $m/2z$ for all the webs but Ythan
  \protect\cite{ks-test1,ks-test2}. The solid lines are the analytical
  predictions of Eq.~(\ref{pred}) for the extremal values of $z$ in the
  empirical data.  Semi-logarithmic plot of the scaled distributions of (c)
  number of prey, and (d) number of predators.  }
%
%

 \label{figure1new}
\end{figure}

Indeed, Eq.~(\ref{prey}) predicts that $P_{{\rm prey}}(k)$ depends only on
$k/2z$. So, we plot in Figs.~\ref{figure1new}a,c the cumulative distributions
$P_{{\rm prey}}(k)$ versus the scaled variable $k/2z$ for the food webs and
find that the data collapse onto a single curve, supporting the possibility
that $P_{{\rm prey}}$ obeys a universal functional form \cite{ks-test1}.

The scaling of $P_{{\rm pred}}(m)$ is not as
straightforward. Equation~(\ref{pred}) indicates that ``true'' scaling holds
only for $m/2z < 1/2$, while for larger values of $m/2z$ there is a Gaussian
decay of the probability function with an explicit dependence on
$z$. However, the decay for $m>2z$ is quite fast and, to first approximation,
not very relevant. Thus, we plot $P_{\rm pred}(m)$ versus the scaled variable
$m/2z$ for the food webs and indeed find a collapse of the data onto a single
curve for $m/2z<0.7$ (Figs.~\ref{figure1new}b,d) \cite{ks-test2}.

\begin{figure}[t!]
 \vspace*{0.cm}
 \includegraphics*[width=7.5cm]{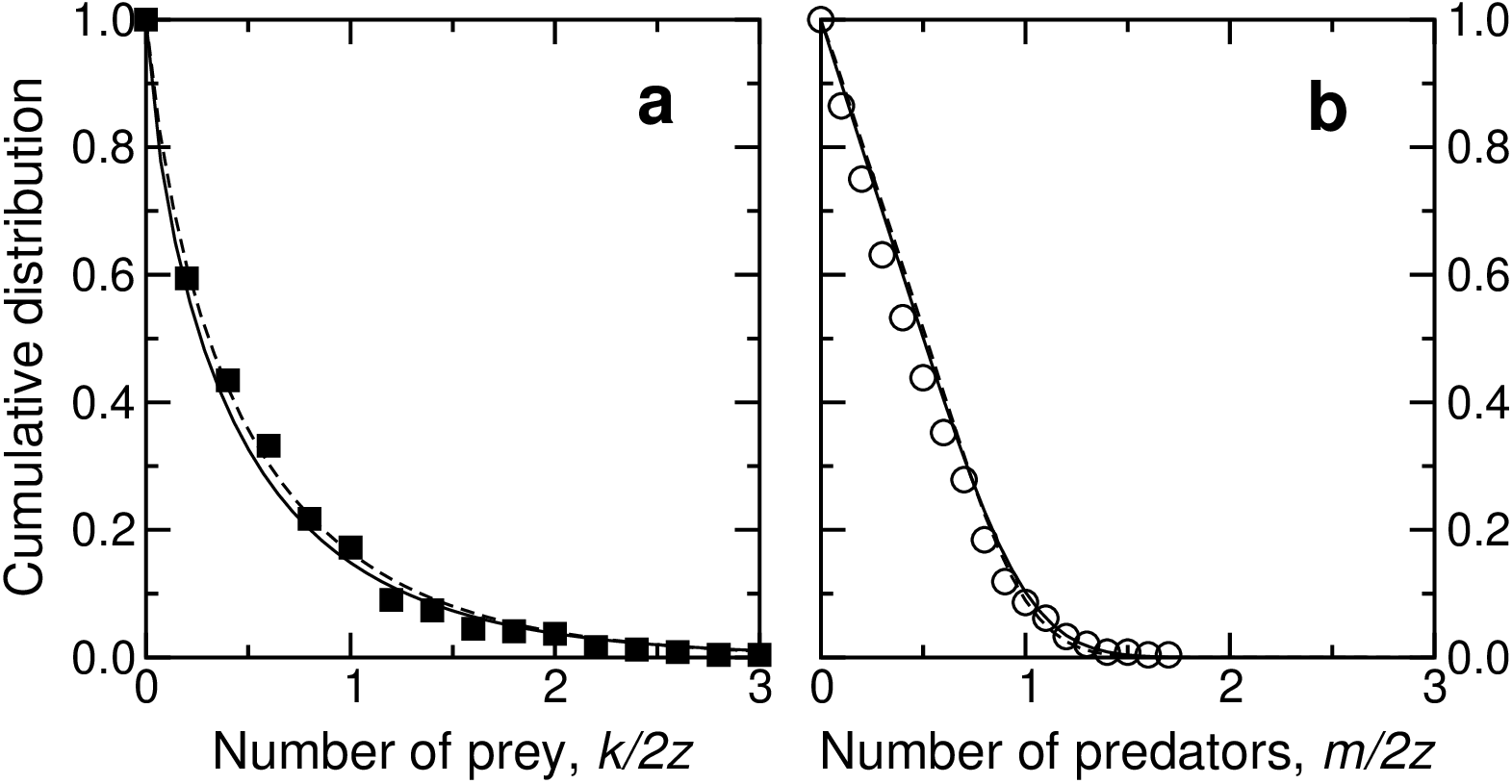}
 \includegraphics*[width=7.5cm]{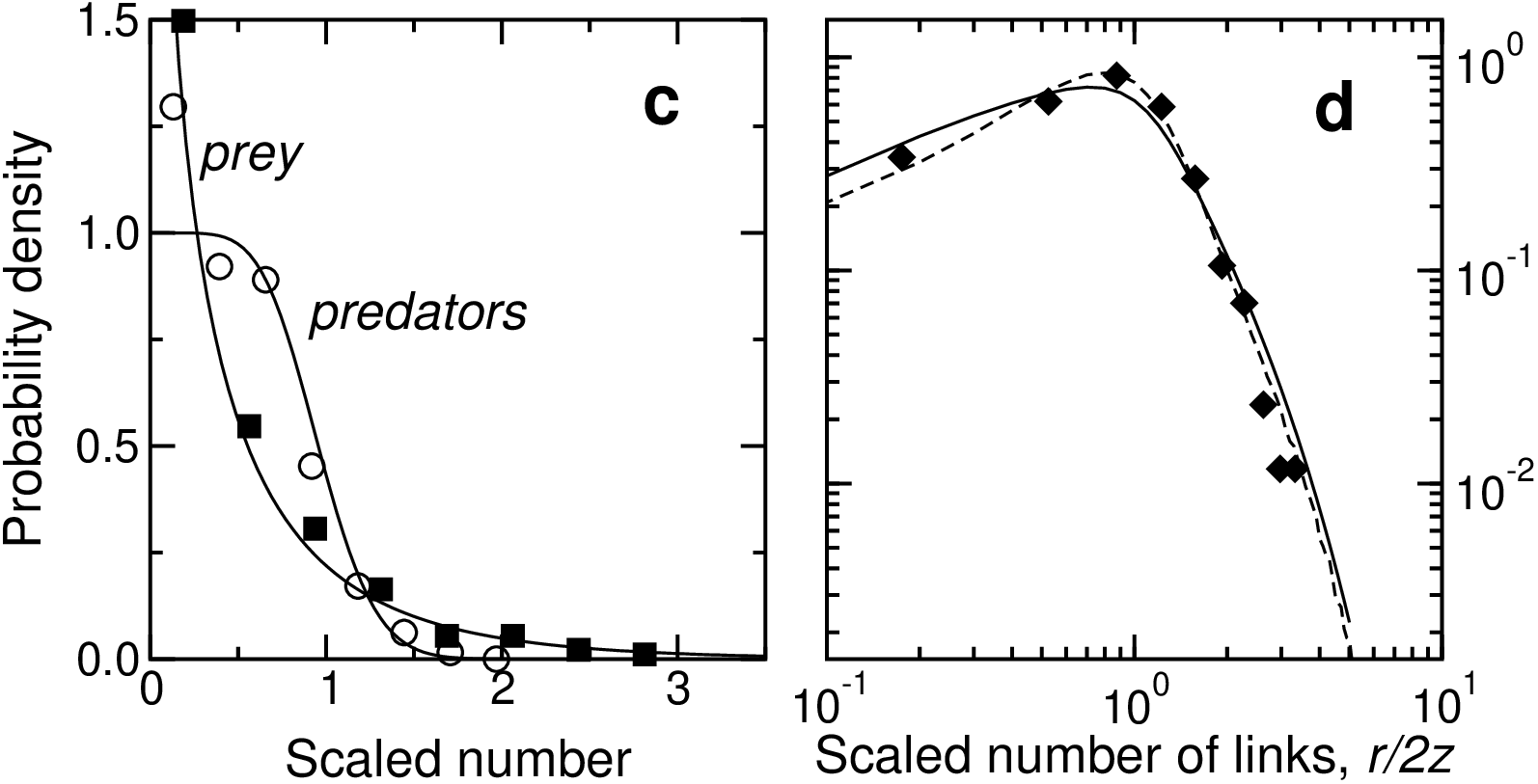}
 \vspace*{-0.3cm}
 \caption{(a) Cumulative distribution $P_{\rm prey}$ of the scaled number of
  prey $k/2z$ for the pooled webs (all except Ythan). The solid line is the
  analytical prediction (\ref{prey}), and the dashed line is a numerical
  simulation of the niche model \protect\cite{Martinez00} with $S=244$ (the
  size of the pooled data) and $z=7.5$ (the average degree for the pooled
  webs).  (b) Cumulative distribution $P_{{\rm pred}}$ of the scaled number
  of predators $m/2z$ for the pooled webs. The solid line is the analytical
  prediction (\ref{pred}) for the case $z=7.5$, and the dashed line is a
  numerical simulation of the niche model \protect\cite{Martinez00} with
  $S=244$ and $z=7.5$.  (c) Comparison of the probability density functions
  of the scaled number of prey and number of predators. It is visually
  apparent that the two distributions have distinct functional forms.  (d)
  Probability density function of the number of trophic interactions per
  species $r=k+m$ pooled for all webs except Ythan. The solid line is
  obtained by numerically convolving the distributions
  (\ref{prey})--(\ref{pred}) while the dashed line is obtained by numerical
  simulations of the niche model in the limit of large web sizes and small
  connectances (we use $S=1000$ and $z = 5$)---the same limit for which the
  analytical curves where derived \protect\cite{Camacho01}.  The tail of the
  distribution decays exponentially, indicating that food webs do {\it not\/}
  have a scale-free structure. }
 \label{figure2}
\end{figure}

Figure~\ref{figure1new} supports the strong new hypothesis that the
distributions of number of prey and number of predators follow {\it universal
functional forms}.  To improve statistics and better determine the specific
functional form of these distributions, one may pool the scaled variables,
$k/2z$ and $m/2z$, from the different webs \cite{ks-test1,ks-test2} into
single distributions, $p_{{\rm prey}}$ and $p_{{\rm pred}} $,
respectively. Figures~\ref{figure2}a,b show the cumulative distributions of
scaled number of prey and scaled number of predators. Note that the
distributions are well approximated by Eqs.~(\ref{prey})--(\ref{pred}) even
though there are no free parameters to fit in the analytical curves. These
results are analogous to the finding of scaling and universality in physical,
chemical and social systems \cite{Bunde94}.
%
%

Figure \ref{figure2}c plots the probability densities for the distribution of
number of prey and number of predators. It is visually apparent that both
distributions are different. This is confirmed by the Kolmogorov-Smirnov test
which rejects the null hypothesis at the $p<0.001$ level. The distribution of
number of prey decays exponentially, and the distribution of number of
predators is essentially a step function with a fast decay.

%
%


One can perform a similar analysis for the distribution $p_{{\rm link}}$ of
the number of trophic links $r\equiv k+m$. As for number of prey or number of
predators, the data from the different webs, upon the scaling $r/2z$,
collapse onto a single curve, further supporting the hypothesis that scaling
holds for food web structure. To better determine the specific functional
form of $p_{\rm link}(r)$, we pool the scaled variables, $r/2z$, from all
webs except Ythan into a single distribution (Fig.~\ref{figure2}d). We find
that $p_{\rm link}(r)$ has an exponential decay for $r/2z\gg 1$, in agreement
with our theoretical calculations. Therefore, there is a characteristic scale
for the linkage density, i.e. food webs do {\it not\/} have a scale-free
structure, in contrast to reports in recent studies of food-web fragility
\cite{Sole}.

%
\begin{figure}[t!]
 \vspace*{0.cm}
 \includegraphics*[width=5cm]{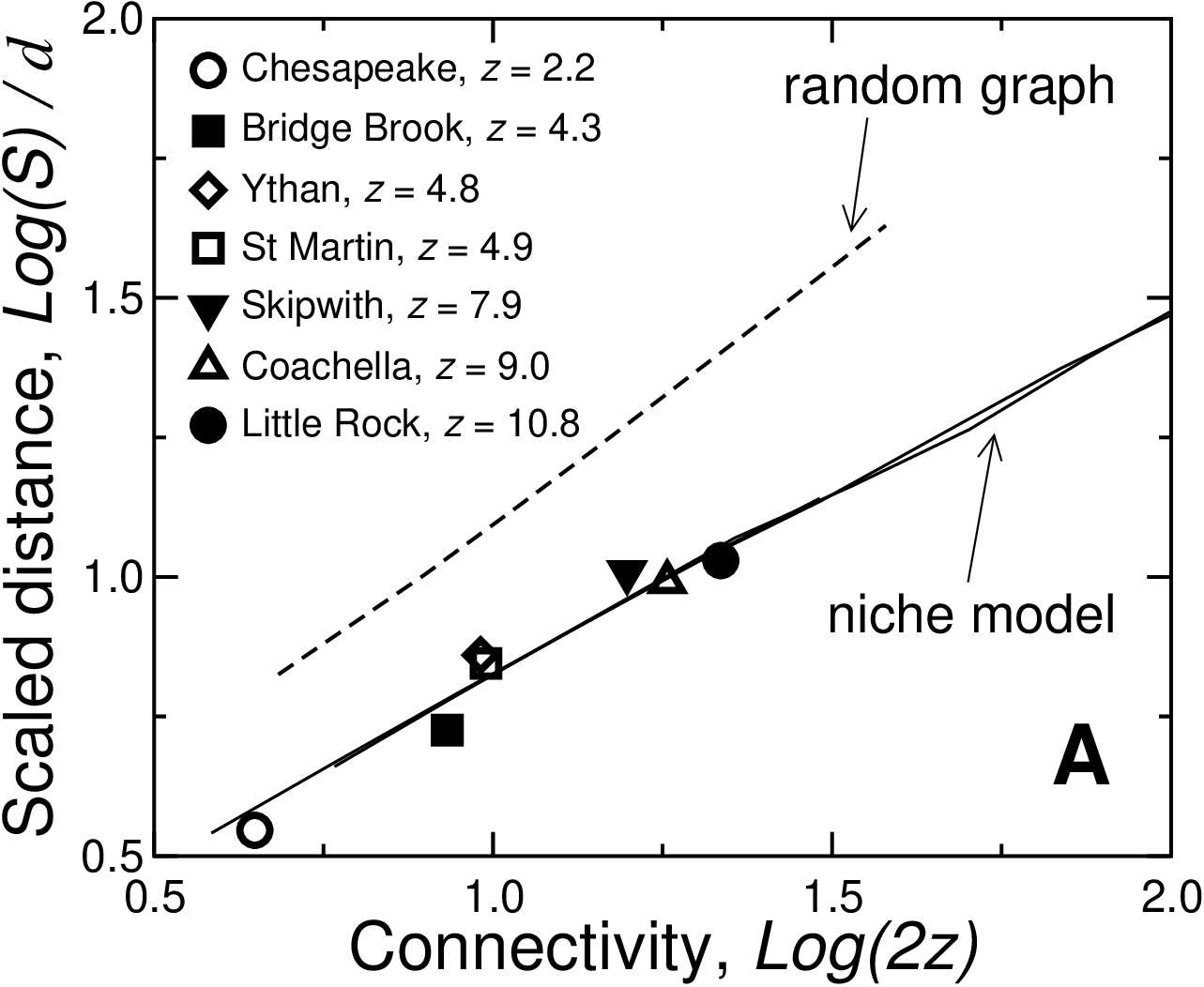}
 \includegraphics*[width=5cm]{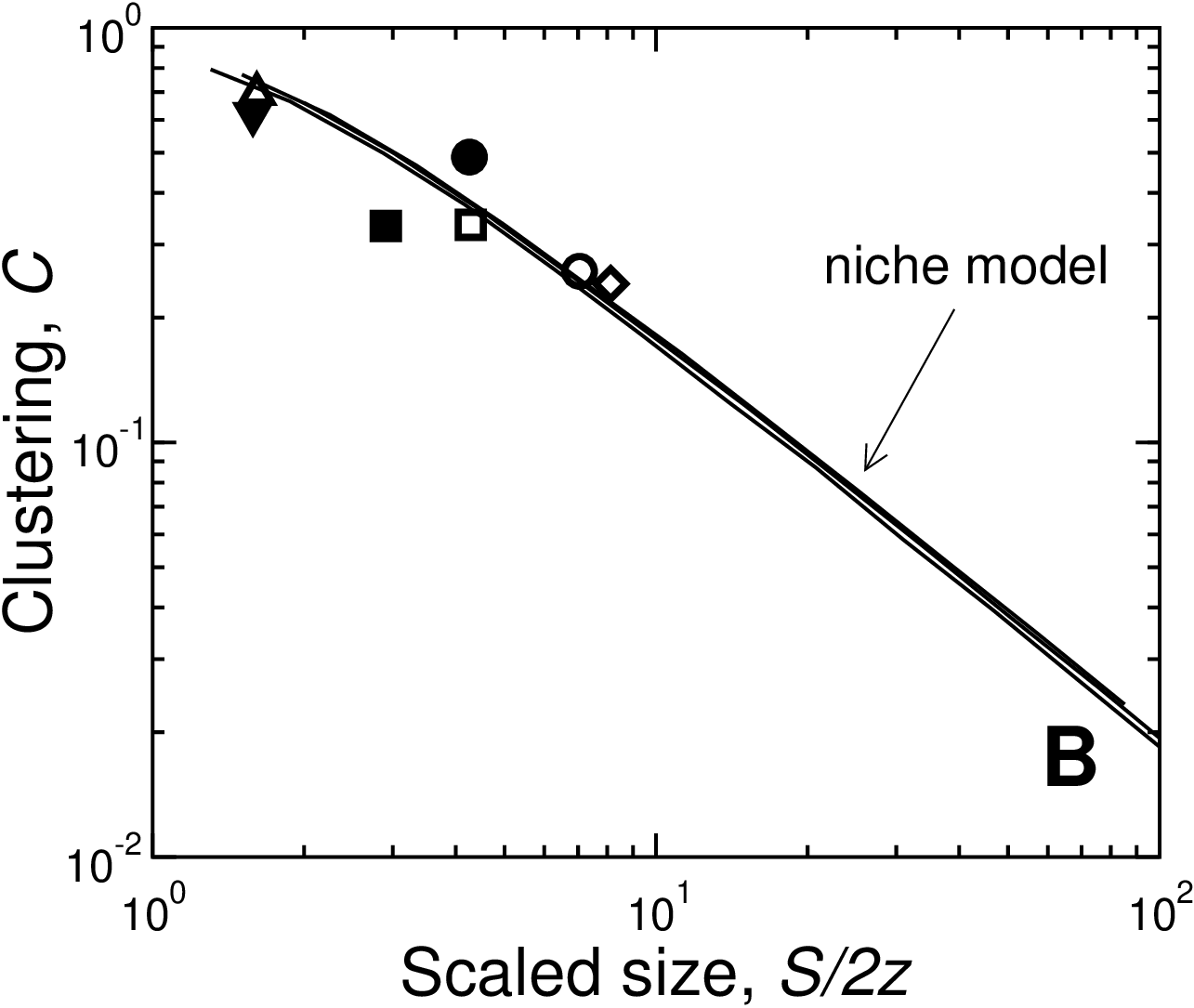}
 \vspace*{-0.3cm}
 \caption{ (a) Scaled average trophic distance $d$ between species versus
  linkage density $z$. We compare the data with the numerical simulations of
  the niche model \protect\cite{Martinez00} for web sizes $S=100, 500, 1000$
  (thin solid lines). We find a logarithmic increase of the average distance
  with web size $S$, in good agreement with the model predictions. We also
  compare our results with the prediction for a random graph with the same
  linkage density as the webs studied (dashed line). The logarithmic
  dependence of $d(S)$ agrees with the expectation for a random graph,
  however, the coefficients of the logarithmic increase differ from the
  predicted values indicating that food webs have a more complex structure
  than that predicted by a random graph. (b) Double-logarithmic plot of the
  clustering coefficient $C$ versus the scaled web size $S/2z$. We compare
  the data with numerical results for the niche model
  \protect\cite{Martinez00} for three values of the linkage density in the
  empirically-relevant range ($z = 2.5, 5, 10$). We find that the clustering
  coefficient of the food webs is inversely proportional to the web size $S$,
  in good agreement with the model predictions and with the asymptotic
  behavior predicted for a random graph \protect\cite{Watts98}. }
 \label{figure3}
\end{figure}
%


Next, we test if the scaling hypothesis suggested by the analysis of
distribution of trophic links also applies to other quantities characterizing
food web structure. We consider two quantities with ecologic implications:
(i) the average trophic distance $d$ between species \cite{Watts98} (which is
the number of species needed to trophically connect two given species), and
(ii) the clustering coefficient $C$ (which counts the fraction of species'
triplets that form fully-connected triangles).  The latter relates to the
compartmentalization in an ecosystem while the former relates to typical
food-chain length.

In Fig.~\ref{figure3}a, we compare our numerical results for the average
trophic distance $d$ for the niche model \cite{Martinez00} with the values
calculated for the food webs analyzed.  We find that $d$ increases with web
size as $\log S$ both for the model and for the data. This logarithmic
increase is the expected behavior for a random graph; however, the slopes
measured for the data and the model are different from the value predicted
for a random graph \cite{Newman01}, suggesting that there is a degree of
``order'' to the connectivity of the food web which may encode the mechanisms
of food web assembly.  Remarkably, this characteristic of the empirical food
webs appears to be captured by the niche model \cite{Martinez00}.  The
results of Fig.~\ref{figure3}a also support the scaling hypothesis and
suggest that the average distance in a food web may also follow a unique
functional form for different food webs.

Figure~\ref{figure3}b shows our results for the clustering $C$ of the food
webs studied and for the niche model \cite{Martinez00}.  We find that the
data is well approximated by the model predictions, and that $C$ decreases to
zero as $1/S$ as web size $S$ increases.


The major finding of this study is the uncovering of unifying quantitative
patterns characterizing the structure of food webs from diverse
environments. Specifically, we find that the distributions of the number of
prey, number of predators and number of links of most of the best studied
food webs seem to collapse onto the same curves after rescaling the number of
links by its average number $z$. Remarkably, the corresponding curves are in
agreement with the analytical predictions of the niche model. Therefore,
these distributions can be theoretically predicted merely by knowing the food
web's linkage density $z$, a parameter readily accessible
empirically. Regularities such as these are interesting as descriptors of
trophic interactions inside communities because they may enable us to make
predictions in the absence of high-quality data, and provide insight into how
communities function and are assembled.

Our results are of interest for a number of other reasons.  First, food webs
do not have a scale free distribution of number of links (total, incoming or
outgoing). This is surprising since one could expect most species to try to
prey on the most abundant species in the ecosystem (an ``abundant-get-eaten''
kind of mechanism).  Such a preferential attachment would lead to a
scale-free distribution of links; instead, we find a single-scale
distribution, suggesting that species specialize and prey on a small set of
other species.
Second, the results of Figs.~\ref{figure3}a,b support the scaling hypothesis
and indicate that there is very little, if any, compartmentalization in
ecosystems
\cite{Pimm80}, suggesting the possibility that ecosystems are highly
interconnected and that the removal of any species may induce large
disturbances.
Third, the structure of food webs is different from many other biological
networks in two important aspects: the links are uni-directional and the in-
and out-degree distribution are different.  These two facts are a result of
the {\it directed\/} character of the trophic interactions and of the
asymmetry it creates.  Interestingly, the niche model captures this asymmetry
in its rules, which may explain its success in explaining the empirical
results.

Our findings are surprising for two reasons: (i) they hold for the most
complete food webs studied, in contrast to previously reported patterns
\cite{ecology}, and (ii) they support the possibility that fundamental
concepts of modern statistical physics such as scaling and universality
\cite{Bunde94}---which were developed for the study of inanimate
systems---may also be applied in the study of food webs---which comprise
animate beings. Indeed, our results are consistent with the underlying
hypothesis of scaling theory \cite{Bunde94}, i.e., food webs display
universal patterns in the way trophic relations are established despite
apparently ``fundamental'' differences in factors such as the environment
(e.g. marine versus terrestrial), ecosystem assembly, and past history.

We thank A. Arenas, J. Bafaluy, M. C. Barbosa, M.  Barth\'el\'emy,
A. D\'{\i}az-Guilera, J. Faraudo, G. Franzese, F. Giralt, S.  Mossa,
R. V. Sol\'e, H. E. Stanley and especially N. D. Martinez and R. J. Williams
for stimulating discussions and helpful suggestions. We also thank
J. E. Cohen, N. D. Martinez and R. J. Williams for making their electronic
databases of food webs available to us. JC and RG thank the Generalitat de
Catalunya and the Spanish CICYT (PPQ2000-1339, BFM2000-0626 and
BFM2000-0351-C03-01) for support. LANA thanks NIH/NCRR (P41 RR13622) and NSF
for support.



\begin{references}

\bibitem{Martinez00}  R. J. Williams and N. D. Martinez,
Nature {\bf 404}, 180
(2000).

\bibitem{Paine92}  R. T. Paine, 
 Nature {\bf 355}, 73
(1992).

\bibitem{McKann98}  K. S. McKann, A. Hastings, and G. R. Huxel, 
Nature {\bf 395}, 794
(1998).

\bibitem{Strogatz01}  S. H. Strogatz 
Nature {\bf 410}, 268
(2001).

\bibitem{Watts98} D. J. Watts and S. H. Strogatz,
Nature {\bf 393}, 440
(1998); M. Barth\'{e}l\'{e}my, and  L. A. N. Amaral,
Phys. Rev. Lett. {\bf 82}, 3180
(1999).

\bibitem{Amaral00}  L. A. N. Amaral,  A. Scala, M. Barth\'{e}l\'{e}my,
and H. E. Stanley,  
Proc. Nat. Ac. Sci USA {\bf 97}, 11149
(2000);
F. Liljeros {\it et al.}
{\it Nature\/} {\bf 411}, 907
(2001).

\bibitem{Newman01} M. E. J. Newman, S. H. Strogatz, and D. J. Watts,
Phys. Rev. E {\bf 64}, 026118 (2001).

\bibitem{ecology} J. E.  Cohen, F. Briand, C. M. Newman,  
{\it Biomathematics\/} {\bf 20} (1990); S. L. Pimm, J. H. Lawton, and
J. E. Cohen,
Nature {\bf 350}, 669
(1991); M. Rejmanek and P. Stary, Nature {\bf 280}, 311 (1979); F. Briand and
J. E. Cohen, Nature {\bf 307}, 264 (1984); {\it ibid}, Science {\bf 238}, 956
(1987).

\bibitem{Lawton88}  J. H. Lawton and P. H. Warren,
Trends Ecol. Evol. {\bf 3}, 242 (1988); T. W. Schoener,
Ecology {\bf 70}, 1559 (1989).

\bibitem{Hall} P. H. Warren, 
Trends Ecol. Evol. {\bf 9}, 136
(1994);
G. A. Polis and D. R. Strong, 
American Naturalist {\bf 147}, 813
(1996);
%
G. A. Polis,
Am. Nat. {\bf 138}, 123
(1991);
%
N. D.  Martinez,
Ecol. Monog. {\bf 61}, 367
(1991);
%
P. H.Warren,  (1989) 
Oikos {\bf 55}, 299
(1989);
%
D. Baird and R. E. Ulanowicz, R. E. (1989) 
Ecol. Monogr. {\bf 59}, 329
(1989).

\bibitem{Hall2}  S. J. Hall and D. Raffaelli,
Adv. Ecol. Res. {\bf 24}, 187
(1993); 
%
H. Huxham, D. Raffaelli, and A. Pike,
J. Anim. Ecol. {\bf 64}, 168 (1995);

\bibitem{Camacho01}  J. Camacho, R. Guimer\`{a}, and L. A. N. Amaral,
Phys. Rev. E {\bf 65} 030901(R) (2002).

\bibitem{numrec}  I. S. Gradstheyn and I.M. Ryzhik, {\it Table of
Integrals, Series and Products}, 6nd ed. (Academic Press, New York, 2000).

\bibitem{smartin}  L. Goldwasser and J. Roughgarden,
Ecology {\bf 74}, 1216 (1993).

\bibitem{ks-test1} To investigate if the species in different webs have
numbers of prey drawn from the same distribution we use the
Kolmogorov-Smirnov test. We find that we cannot reject the null hypothesis
that the species for all webs have number of preys drawn from the same
underlying distributions.  The case of Little Rock is marginal with regard to
the Kolmogorov-Smirnov test but visual inspection suggests that this web follows the same patterns as the other.

\bibitem{ks-test2} The Kolmogorov-Smirnov test shows that we cannot reject
the null hypothesis that the species for all webs have number of predators
drawn from the same underlying distributions except for the case of Ythan
\protect\cite{Hall91}. For the Ythan Estuary web, we find results consistent
with an exponential distribution of number of predators. The finding that
Ythan is different from other webs may be explained in two ways: (i) the Ythan
Estuary web appears to be still quite incomplete for bottom and top species
\protect\cite{Martinez00,Hall2}, or (ii) Ythan belongs to a different
universality class.

\bibitem{Hall91} S. J.  Hall and D. Raffaelli,
J. Anim. Ecol. {\bf 60}, 823
(1991).

\bibitem{Bunde94}  Bunde, A., Havlin S. (eds.) (1994) {\it Fractals in
Science\/} (World Scientific, Singapore).

\bibitem{Sole} R. V. Sol\'{e} and J. M.  Montoya,
Proc. R. Soc. B. {\bf 268}, 2039
(2001);
J. M. Montoya and R. V. Sol\'{e}, 
J. Theor. Biol. (2001).

\bibitem{Pimm80}  S. L. Pimm, and J. H. Lawton
J. Animal Ecol. {\bf 49}, 879
(1980).

\end{references}
\end{document}